\documentclass[aps,reprint,groupedaddress]{revtex4-1}
\usepackage{graphicx}
\usepackage{dcolumn}
\usepackage{bm}
\usepackage{color}
\usepackage{array}
\usepackage{mathrsfs}
\usepackage{ulem} 
\usepackage{amsmath}
\usepackage{amssymb}
\usepackage{url}

\begin{document}


\title{Magnetization plateau and supersolid phases in the spin-1/2 antiferromagnetic Heisenberg model on a tetragonally distorted fcc lattice}

\author{Katsuhiro Morita}
\email[e-mail:]{katsuhiro.morita@rs.tus.ac.jp}
\affiliation{Department of Applied Physics, Tokyo University of Science, Tokyo 125-8585, Japan}

\author{Takami Tohyama}
\affiliation{Department of Applied Physics, Tokyo University of Science, Tokyo 125-8585, Japan}

\date{\today}

\begin{abstract}
The Heisenberg model on a face center cubic (fcc) lattice is a typical three-dimensional frustrated spin system expected to have magnetization plateaus and supersolid phases.
There are model compounds $A_2$CoTeO$_6$ ($A$ = Ca, Sr, Pb) for the fcc lattice but with lattice distortions.
Motivated by the presence of the compounds, we investigate the ground state of the spin-1/2 antiferromagnetic Heisenberg model on a tetragonally distorted fcc lattice in the magnetic field using a large-size cluster mean-field method for the sake of finding new supersolid phases. We find five supersolid phases in the model, indicating possibility to observe supersolid phases in these compounds. We also find that one of the supersolid phases is similar to the nonclassical coplanar phase obtained in the XXZ model on the triangular lattice.
\end{abstract}

\maketitle

\section{INTRODUCTION}
\label{Sec1}
Highly frustrated strongly correlated systems have been studied for decades because of a good playground to find exotic quantum states~\cite{Frus1,Frus2,Frus3}. One of the playgrounds is the magnetization process of frustrated quantum spin systems, where magnetization plateaus induced by quantum fluctuations have been studied both theoretically and experimentally~\cite{kagothe1,kagothe2,kagothe3,kagothe4,kagothe5,kagoexp1,kagoexp2,kagoexp3,J1-J21,J1-J22,J1-J23,che1,che2,trithe1,trithe2,trithe4,trithe5,trithe6,trithe7,trithe8,trithe9,triex1,triex2,oexp1,oexp2,oexp3,oexp4,oexp5,oexp6}.
For example, a 1/3  magnetization plateau has been observed in the triangular lattice (TL) compound Ba$_3$CoSb$_2$O$_9$ described as a XXZ quantum spin system with weak easy-plane anisotropy~\cite{triex1,triex2}. It is interesting that in the same XXZ model on the TL but with large easy-plane anisotropy, a nonclassical coplanar state induced by quantum fluctuations has been proposed theoretically~\cite{trithe4,xxztri}. The coplanar state in this case is called supersolid (SS) phase.

In condensed matter physics, studies finding SS phases have been carried out repeatedly~\cite{SSHe}.
In quantum spin systems, a solid state can be defined as a state with a diagonal long-range order such as an up-up-down state corresponding to the 1/3 magnetization plateau in the TL.
On the other hand, a superfluid (SF) state is defined as a state with an off-diagonal long-range order that breaks $U(1)$ symmetry.
Accordingly a SS state is defined as a state having both the orders of the solid and the SF phases~\cite{SSspin}.
The search for SS phases has extensively been performed in recent years~\cite{SS1,SS2,SS3,SS4,SCBO1,SCBO2,Crsp2,Crsp3}. For example, the SS phases have been observed in SrCu$_2$(BO$_3$)$_2$ with spin-1/2 Shastry-Sutherland lattice~\cite{SCBO1,SCBO2} and Cr spinel compounds with spin-3/2 pyrochlore lattice~\cite{Crsp2,Crsp3}.

Finding new SS phases is an important issue for fully understanding frustrated quantum spin systems.
One of unexploited systems is the Heisenberg model on a face center cubic lattice (fccL), which is a typical three-dimensional frustrated spin system expected to have magnetization plateaus and SS phases. 
There are model compounds for the fccL, which are $A_2$CoTeO$_6$ ($A$ = Ca, Sr, Pb) with the double perovskite structure~\cite{CFFcry1,CFFcry2,CFFcry3}. 
The fccL compounds exhibit lattice distortions that may induce new phases due to strong frustration.
Therefore, it is necessary to investigate theoretically the ground state of the distorted fccL for the sake of finding new phases such as SS phases and to provide helpful hints for forthcoming experiments.

In this paper, we investigate the ground states of the spin-1/2 Heisenberg model on a tetragonally distorted fccL (TDfccL) at zero temperature in a magnetic filed using a large-size cluster mean-field (CMF) method.
We find eight phases: a solid, two SF, and five SS phases. The solid phase has up-up-up-down structure corresponding to a 1/2 magnetization plateau.
One of the SS states at high magnetic field is similar to the nonclassical coplanar state obtained in the XXZ model on the TL~\cite{trithe4,xxztri}.

This paper is organized as follows. The Heisenberg model on the TDfccL is introduced in Sec.~\ref{Sec2}. In Sec.~\ref{Sec3}, the CMF method is described. In Sec.~\ref{Sec4}, the magnetization process and the magnetic phase diagram are shown, and the magnetic structures and the symmetry breaking would be discussed.
We discuss the relationship between the new phases obtained in the present study and those in the previous study in Sec.~\ref{Sec5}.
 Finally, a summary is given in Sec.~\ref{Sec6}.
\begin{figure}[tb]
 \begin{center}
	\includegraphics[width=40mm]{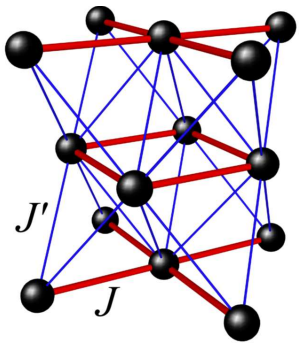}
   \caption{Lattice structure of TDfccL.  The red solid and blue thin lines denote the exchange interactions $J$ and $J^\prime$, respectively. }
  \label{model}
\end{center}
\end{figure}

\section{Model}
\label{Sec2}
The Hamiltonian for the spin-1/2 TDfccL in the magnetic field is given by
\begin{eqnarray} 
H &=& J\sum_{\langle i,j \rangle } \mathbf{S}_i \cdot \mathbf{S}_j +J^\prime\sum_{\langle\langle i,j \rangle\rangle } \mathbf{S}_i \cdot \mathbf{S}_j - h\sum_i S^{z}_i,
\label{hami}
\end{eqnarray}
where $\mathbf{S}_i$ is the spin-1/2 operator at site $i$, $\langle i,j \rangle$ and $\langle\langle i,j \rangle\rangle$ runs over the nearest-neighbor spin pairs with exchange interaction $J$ and $J'$ shown in Fig.~\ref{model}, respectively, and $h$ is the magnitude of the magnetic field in the z-direction.
Since the TDfccL can be divided into tetrahedra,  the Hamiltonian (\ref{hami}) may be rewritten by the sum of partial Hamiltonians of the tetrahedra as $H = \sum_{t} h^{(t)}_\mathrm{tetra},$
where $t$ means an index of the tetrahedra and the summation is performed for all the tetrahedra. 
The Hamiltonian of a tetrahedron, $h^{(t)}_{tetra}$, is defined by
\begin{eqnarray} 
h^{(t)}_\mathrm{tetra} &=& \frac{J}{2} (\mathbf{S}_1^{(t)} \cdot \mathbf{S}_2^{(t)} + \mathbf{S}_3^{(t)} \cdot \mathbf{S}_4^{(t)} ) \nonumber \\
&+& \frac{J^\prime}{2}  [\mathbf{S}_1^{(t)} \cdot (\mathbf{S}_3^{(t)} + \mathbf{S}_4^{(t)}) + \mathbf{S}_2^{(t)} \cdot (\mathbf{S}_3^{(t)} + \mathbf{S}_4^{(t)})  ] \nonumber \\
&-& \frac{h}{4} (S^{z(t)}_1+S^{z(t)}_2+S^{z(t)}_3+S^{z(t)}_4),
\label{localhami}
\end{eqnarray}
where $i$ (=~1, 2, 3, and 4) in $\mathbf{S}_i^{(t)}$ represents the site number in the $t$-th tetrahedron and the factors of 1/2 (1/4) in the exchange (Zeeman) terms are due to double (quadruple) counting.
We note that four sublattice magnetic orders are implicitly assumed in rewriting the Hamiltonian (\ref{hami}) by the sum of the partial Hamiltonians (\ref{localhami}).

\section{cluster mean-field method}
\label{Sec3}
\begin{figure}[tb]
 \begin{center}
	\includegraphics[width=80mm]{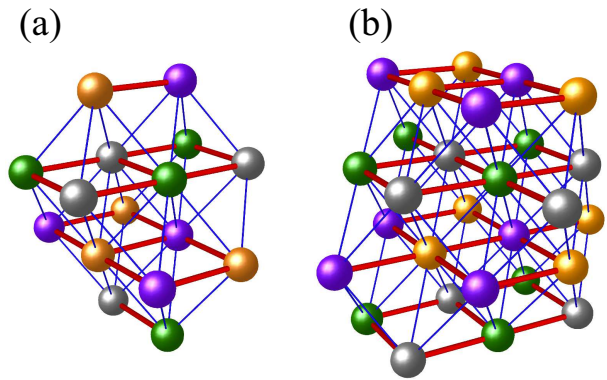}
   \caption{Four-sublattice structure with orange, purple, green, and gray sublattices used in our CMF method for the TDfccL. (a) $N=16$ and (b) $N=28$.
 The red solid and blue thin lines denote the exchange interactions $J$ and $J^\prime$, respectively. }
  \label{CMF}
\end{center}
\end{figure}

The CMF method has been successfully applied to the analysis of the frustrated spin models~\cite{CMF1,CMF2,CMF3,CMF4}.
Since a mean-field approximation gives an exact solution in infinite dimensions, the CMF method based on a mean-field approximation is expected to be suitable for the three-dimensional TDfccL with magnetic orders.
Therefore, employ the CMF method used in \cite{trithe6} and perform CMF calculations at zero temperature to find the ground states of Hamiltonian (\ref{hami}). We use two clusters with the number of spin $N=16$ and 28, constituting four-sublattice structure as shown in Fig.~\ref{CMF}, where four colors on the spheres correspond to the four sublattices. 
We note that the two clusters are isotropic with respect to the $a$, $b$, and $c$ axes if $J=J'$~\cite{tetra}.

In order to determine the first order transition points, we need to obtain the energy difference between different spin configurations.
At magnetic field $h$, the energy $E(h)$ is given by integrating the magnetization $M(h)$ with respect to $h$:
\begin{eqnarray}
E(h) = E(h_s) + \int_h^{h_s} M(h') dh',
\end{eqnarray}
where $h_s$ is the magnetic field giving rise to the saturated magnetization $M_\mathrm{sat}$ and $E(h_s)$ is easily obtained.

\section{results}
\label{Sec4}

\begin{figure}[tb]
 \begin{center}
	\includegraphics[width=86mm]{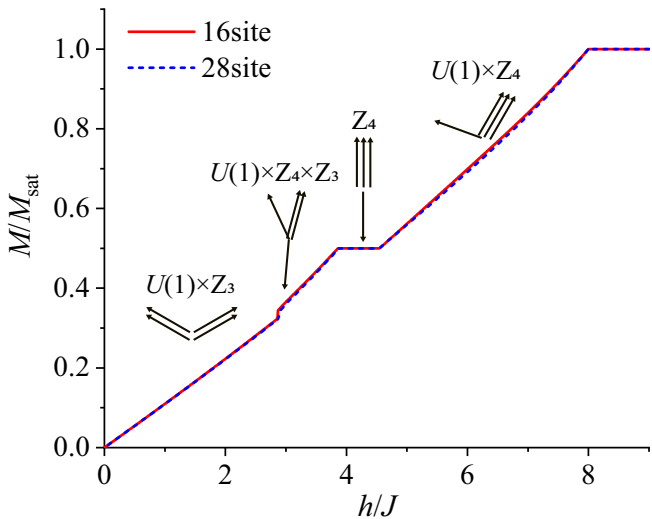}
   \caption{Magnetization curve of the fccL ($J'=J$) at zero temperature using the CMF method with $N=16$ (red solid line) and $N=28$ (blue dashed line) clusters. 
The four arrows on each region indicate schematic magnetic structure in the four sublattices, where the upward direction corresponds to the direction of the magnetic field.
The symbols above the arrows indicate the broken symmetry in each structure.
Each phase corresponds to SF, SS, solid, and SS phases from small $h$ to large $h$.}
  \label{M-H}
\end{center}
\end{figure}

We first show the magnetization curve at $J=J^\prime$, which is shown in Fig.~\ref{M-H}.
Since there is little difference of the magnetization curve between $N=16$ and $N=28$, the finite size effect is expected to be small. 
We obtain four phases except for the full moment (saturated magnetization) phase. 
The arrows and the symbols in Fig~\ref{M-H} represent schematic magnetic structures in the four sublattices and the broken symmetry in each phase, respectively.
The symmetry $\mathbb{Z}_4$ ($\mathbb{Z}_3$) corresponds to the degrees of freedom choosing one from four (three) spins, and the $U(1)$ corresponds to the rotation symmetry in the direction of the magnetic field. These symmetry notations also describe the degeneracy of the ground state.
All magnetic structures become coplanar or collinear structure generated by so-called ``order-by-disorder'' mechanism. 
Judging from the broken symmetry and magnetic structure, we can assign each phase as SF, SS, solid, and SS phases in the order of increasing $h$.
In Fig.~\ref{M-H}, a first-order transition with a spin-flop occurs at $h/J\approx3$ before appearing the 1/2 magnetization plateau at $M/M_\mathrm{sat}=1/2$.
This property is similar to that obtained in the distorted TL~\cite{J1-J22}, $J_1$-$J_2$ square lattice~\cite{J1-J23}, and pyrochlore lattice~\cite{Crsp2,Crsp3}.
We note that the magnetic structures and the first-order transition in Fig.~\ref{M-H} are the same as those reported previously~\cite{fccthe}.
These agreements encourage the use of the CMF method for the analysis of the TDfccL.

\begin{figure}[tb]
 \begin{center}
	\includegraphics[width=86mm]{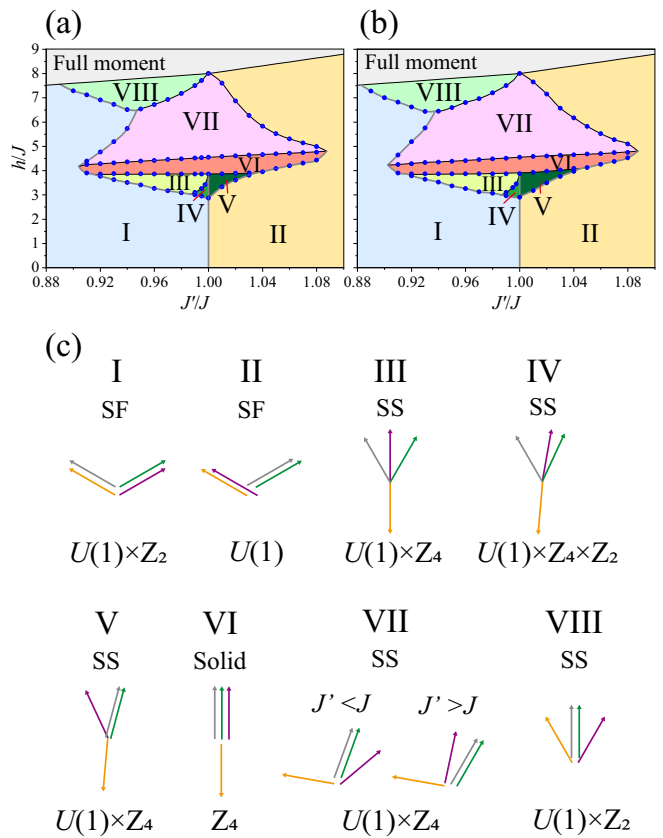}
   \caption{Phase diagram of the TDfccL at zero temperature in a magnetic field using the CMF method with (a) $N=16$ and (b) $N=28$ clusters.
The blue dots denote the phase transition points determined by the CMF method. 
The solid gray and thin black lines denote first-order and second-order phase boundaries, respectively.
The number from I to VIII represent different phases.
(c) Schematic magnetic structures in the I--VIII phases, where the upward direction corresponds to the direction of the magnetic field. The colors of the arrows correspond to those of the spheres shown in Fig.~\ref{CMF}. 
The characters under the arrows indicate the broken symmetry in each phase.
The I and II phases are SF, the VI phase (the 1/2 plateau phase) is solid, and the III, IV, V, VII, and VIII phases are SS. }
  \label{PH}
\end{center}
\end{figure}

Next, we investigate the ground states of the TDfccL. 
Figures~\ref{PH}(a) and~\ref{PH}(b) show phase diagrams of the TDfccL at zero temperature in the plane of $J^\prime$ and $h$ for the $N=16$ and $N=28$ clusters, respectively.
The blue dots in Figs.~\ref{PH}(a) and~\ref{PH}(b) denote the phase transition points determined by the CMF method.
The solid gray (thin black) lines denote first-order (second-order) phase boundaries expected from the calculated phase transition points.
In order to obtain a more accurate phase diagram in the thermodynamic limit, it would be necessary to use the scaling scheme~\cite{trithe6}. However, there is little difference between Figs.~\ref{PH}(a) and~\ref{PH}(b), indicating small size effects.
We find eight phases from I to VIII, all of whose magnetic structure become coplanar or collinear structure as shown in Fig.~\ref{PH}(c), where the four arrows on each phase indicate schematic magnetic structure in the four sublattices.
In the classical spin system, the ground states become only I or II spin structures except for $J'=J$ (not shown here).
Therefore, the III--VIII phases are induced by quantum fluctuations.

We here mention the characteristics of these phases.
The broken symmetry in each phase is shown as the symbols below the arrows in Fig.~\ref{PH}(c).
The $\mathbb{Z}_4$ in the III--VII phases corresponds to the degree of freedom choosing a downward or leftward spin [colored by orange in Fig.~\ref{PH}(c)] from four spins.
The $\mathbb{Z}_2$ in the I and IV phases corresponds to the degree of freedom exchanging the green spin for gray spin.
The $\mathbb{Z}_2$ in the VIII phase corresponds to the degree of freedom exchanging the green and gray spin pair for orange and purple spin pair.
The $U(1)$ corresponds to the rotation symmetry in the direction of the magnetic field.
The I and II phases belong to SF, which is defined as the state with magnetic order in a plane perpendicular to the $z$ direction and with uniform local magnetization $\langle S_i^z\rangle$.
The VI phase being the 1/2 plateau phase with up-up-up-down structure belongs to solid, which has a magnetic order in parallel to the $z$ direction.
The III, IV, V, VII, and VIII phases belong to SS having features of both SF and solid. 
We note that the III, IV, V, VII, and VIII phases cannot be described by the linear spin-wave (LSW) method because these phases do not satisfy the required condition for the LSW method that the ground state should be on a saddle point of the energy landscape~\cite{J1-J22}. Therefore, we consider that the CMF used in the present study is one of the best methods for describing the frustrated Heisenberg models.

\section{DISCUSSIONS}
\label{Sec5} 

We compare our results on the TDfccL with the known results of the XXZ model with large easy-plane anisotropy on the TL~\cite{trithe4,xxztri}.
The magnetic structure in the VIII phase, i.e., the SS phase, is similar to the structure of the nonclassical coplanar (so-called $\pi$-coplanar or $\Psi$) state obtained on the TL~\cite{trithe4,xxztri}. 
In fact, the $\pi$-coplanar structure is obtained by removing either green or gray spin in the VIII phase [see Fig.~\ref{PH}(c)].
We should emphasize that the VIII phase is obtained in the Heisenberg model without strong anisotropy in contrast to the $\pi$-coplanar phase. Therefore, our finding of the SS phases in the TDfccL may encourage efforts to observe the SS states in spin-1/2 systems.

As seen in Fig.~\ref{PH}(c), the green and gray spins in the spin structure of the VIII phase align completely their direction to the magnetic field. This alignment is continuously achieved from the spin structure of the VII phase when $h$ increases. This means that the VII $\rightarrow$ VIII phase transition is a continuous one, i.e., second-order transition.
Experimentally,  a phase transition similar to the VII $\rightarrow$ VIII phase transition has been observed in Cr spinel compounds with the spin-3/2 pyrochlore lattice at the high magnetic field, i.e., ZnCr$_2$O$_4$ at 350~T~\cite{Crsp3} and HgCr$_2$O$_4$ at 36~T~\cite{Crsp1}, although the magnetic structure of the phases have not been specified.
As for the present the VII $\rightarrow$ VIII phase transition, the phase transition in the Cr spinel compounds cannot be explained by the analysis of a classical spin system~\cite{Crthe1,Crthe2}.
Although there are differences in spin size and lattice geometry between the Cr spinel lattice and our TDfccL, 
there are similarities each other in terms of three-dimensional frustrated lattice and four sublattice structure.
Therefore, we believe that the VII $\rightarrow$ VIII transition at the high magnetic field obtained in our TDfccL might be related to the phase transitions of the Cr spinel compounds and the phases at higher magnetic field might have the same characteristic of the VIII phase.

We suppose that other interactions which the Co compounds with the fccL can have, such as anisotropic interactions and second neighbor interactions, may induce new phases not obtained in the present study. Therefore, the synthesis of new fcc magnets is highly desired to get an insight into the new phases not discussed yet.

\section{summary}
\label{Sec6} 
Inspired by the recent understanding of the SS phase in frustrated quantum spin systems and the presence of highly frustrated fccL compounds, 
we investigated the ground state of the spin-1/2 Heisenberg model on the TDfccL in the magnetic field using the large-size CMF method.
We obtained phase diagrams at zero temperature and found five SS phases.
The magnetic structure in one of these phases is similar to the $\pi$-coplanar state induced by quantum fluctuation in the TL with large easy-plane anisotropy.
We believe that our results are closely related to the phase transition that cannot be explained by the classical spin system observed with Cr spinel compounds.
Our study will motivate new experimental investigations on fccL compounds in the future.
\begin{acknowledgments}
We thank H. Tanaka for useful discussions. 
This work was supported by MEXT, Japan, as a social and scientific priority issue (creation of new functional devices and high-performance materials to support next-generation industries) to be tackled by using a post-K computer. The numerical calculation was carried out at the facilities of the Supercomputer Center, the Institute for Solid State Physics, the University of Tokyo.
\end{acknowledgments}


\begin{thebibliography}{00}
\bibitem{tetra} We note that we cannot use a four-site cluster in our CMF method because of the presence of anti-Neel points.
\end{thebibliography}
\end{document}